\documentclass[conference]{IEEEtran}

\IEEEoverridecommandlockouts
\usepackage{cite}
\usepackage{amsmath,amssymb,amsfonts}
\usepackage{algorithmic}
\usepackage{graphicx}
\usepackage{textcomp}
\usepackage{xcolor}

\usepackage[linesnumbered,ruled]{algorithm2e} 
\usepackage[capitalize]{cleveref}

\usepackage{amsthm}

\usepackage{caption}
\usepackage{subcaption}

\crefalias{AlgoLine}{line}
\Crefname{algocfline}{line}{lines}
\Crefname{algocf}{Algorithm}{Algorithms}
\Crefname{AlgoLine}{Line}{Lines}
\Crefname{fig}{Figure}{Figures}

\DeclareMathOperator*{\argmax}{argmax}

\newtheorem{theorem}{Theorem}
\newtheorem{lemma}{Lemma}
\newtheorem{definition}{Definition}
\newtheorem*{assumption}{Assumption}
\newtheorem*{task selection}{Method}
\def\BibTeX{{\rm B\kern-.05em{\sc i\kern-.025em b}\kern-.08em
    T\kern-.1667em\lower.7ex\hbox{E}\kern-.125emX}}
\begin{document}

\title{Partitioned Scheduling for DAG Tasks 
\\Considering Probabilistic Execution Time\\

\thanks{This work was supported by JST AIP Acceleration Research \textbf{JPMJCR25U1} and JST CREST Grant Number \textbf{JPMJCR23M1}, Japan.}
}



\author{
    \IEEEauthorblockN{Fuma Omori\IEEEauthorrefmark{1}, Atsushi Yano\IEEEauthorrefmark{1}\IEEEauthorrefmark{2} and Takuya Azumi\IEEEauthorrefmark{1}\IEEEauthorrefmark{2}, }
    \IEEEauthorblockA{\IEEEauthorrefmark{1}
        \textit{Graduate School of} 
        \textit{Science and Engineering,}
        \textit{Saitama University, Japan}
    }
    \IEEEauthorblockA{\IEEEauthorrefmark{2}
        \textit{TIER IV Incorporated, Japan}
    }
}

\maketitle
\begin{abstract}
Autonomous driving systems, critical for safety, require real-time guarantees and can be modeled as DAGs. Their acceleration features, such as caches and pipelining, often result in execution times below the worst-case.
%
Thus, a probabilistic approach ensuring constraint satisfaction within a probability threshold is more suitable than worst-case guarantees for these systems.
This paper considers probabilistic guarantees for DAG tasks by utilizing the results of probabilistic guarantees for single processors, which have been relatively more advanced than those for multi-core processors. 
This paper proposes a task set partitioning method that guarantees schedulability under the partitioned scheduling.
The evaluation on randomly generated DAG task sets demonstrates that the proposed method schedules more task sets with a smaller mean analysis time compared to existing probabilistic schedulability analysis for DAGs.
The evaluation also compares four bin-packing heuristics, revealing Item-Centric Worst-Fit-Decreasing schedules the most task sets.

\end{abstract}

\section{Introduction}
Autonomous driving systems (e.g., Waymo Driver~\cite{waymo}, Autoware~\cite{autoware}) hold promise for reducing traffic accidents, congestion, and logistics issues. Guaranteeing their real-time performance requires ensuring system processing completion within deadlines, making accurate response time analysis vital.

Modeling the complex execution scenarios of autonomous driving systems, arising from sensor data, multiple processes, and their interdependencies, is challenging. To address this, directed acyclic graphs (DAGs)~\cite{AV-DAG3--DAG-WCET1,AV-DAG1,AV-DAG2} are employed, where system processes are nodes and dependencies are directed edges, simplifying response time analysis.

Existing research on DAG task systems in hard real-time systems analyzes response times based on the worst-case execution time (WCET) of each node~\cite{DAG-WCET2,AV-DAG3--DAG-WCET1}.
However, modern autonomous driving hardware (e.g., caches, pipelining) often executes tasks significantly faster, leading to overly pessimistic WCET-based response times that impede practical system design.
This issue is addressed by real-time scheduling approaches that model node execution time as a probability distribution based on actual statistics and provide probabilistic guarantees of deadline satisfaction.

Probabilistic Worst-Case Execution Time (pWCET)~\cite{pWCET} safely over-approximates execution time probabilistically. Its common use stems from allowing the IID probability assumption, treating process execution times independently despite dependencies such as caching. pWCET distributions are derived via Static Probabilistic Timing Analysis (SPTA) and Measurement-Based Probabilistic Timing Analysis (MBPTA)~\cite{PTA-hybrid,probability-timing-survey,MBPTA2022}. This study assumes given pWCET distributions without addressing their derivation.


Few probabilistic response time methods~\cite{DAGprob,SHPC-DAG,DAG-E2E} exist for DAG tasks with pWCET or probabilistic execution times. 
However, in general, probabilistic analysis methods consider multiple states; thus, the analysis methods are restricted to small task sets. 
Compared to DAG task analysis, single-core probabilistic analysis~\cite{MonteCarlo,probability-survey,berry-essen,EDF,2023-FP-RT} is more developed, with techniques for reducing analysis time.
%

This paper proposes a DAG task set partitioning method
under the partitioned Earliest Deadline First (EDF) scheduling. 
Each node of the DAG task has a probabilistic execution time. 
To summarize, this paper makes the following contributions: 

\begin{itemize}
    \item 
    By considering probabilistic execution time, the proposed partitioning method schedules task sets exceeding a worst-case utilization of one, unlike conventional WCET-based methods. This increases schedulability and reduces pessimism, improving system efficiency. 
    \item 
    Treating DAG tasks as single-node tasks under EDF, the proposed method outperforms existing schedulability analysis in the number of scheduled task sets and the mean analysis time, enabling analysis of previously intractable task sets.
    \item 
    Adopting an Item-Centric Worst-Fit-Decreasing strategy for task set partitioning enables the proposed method to guarantee schedulability in systems with higher utilization compared to other bin-packing strategies.
\end{itemize}

The remainder of this paper is organized as follows. 
\cref{sec: sys} describes the system model of this paper.
\cref{sec: proposed approch} introduces the proposed method. 
\cref{sec: app2fed} describes the application of the proposed method to federated scheduling.
\cref{sec: evaluation} evaluates and discusses the proposed method.
\cref{sec: related-work} introduces related work.
\cref{sec: concl} presents the conclusion and future work.

\section{System Model}
\label{sec: sys}
In this section, the system model is described. 
Notations in this paper are listed in~\cref{tab:symbol}. 
\begin{table}[t]
    \centering
    \caption{Summary of notations}
    \begin{tabular}{|l|p{4.7cm}|}
        \hline
        \textbf{Symbol} & \textbf{Description}\\ 
        \hline
         $\Gamma = \{\tau_1,...,\tau_n\}$ & Task set to be scheduled\\
         \hline
         $\tau_i=(G_i, T_i, D_i, \rho_i)$ & DAG task\\
         \hline
         $G_i=(V_i, E_i)$& DAG\\
         \hline
         $T_i$& Period of $\tau_i$\\
         \hline
         $D_i$& Relative deadline of $\tau_i$\\
         \hline
         $\rho_i$& Threshold specifying the maximum acceptable deadline failure probability for $\tau_i$\\
         \hline
         $V_i$& Set of nodes of DAG $G_i$\\
         \hline
         $E_i$& Set of edges of DAG $G_i$\\
         \hline
         $v_{i,j} \in V_i \ ( j\in \{1,...,|V_i|\})$ & Node (subtask) of DAG $G_i$\\
         \hline
         $(v_{i,j}, v_{i,k}) \in E_i$& Edge of DAG $G_i$ from node $v_{i,j}$ to $v_{i,k}$\\
         \hline
         $C_{i,j}$& pWCET distribution related to node $v_{i,j}$\\
         \hline
         $C_i$ & Probabilistic total execution time distribution defined in~\cref{def: PTETD}\\
         \hline
         $J_{i,j}$ & $j$-th job released from $\tau_i=(C_i, T_i, D_i, \rho_i)$\\
         \hline
         $d_{i,j}$ & Absolute deadline of $J_{i,j}$\\
         \hline
         $C_i'$ & Adjusted probabilistic total execution time distribution defined in~\cref{def: APTETD}\\
         \hline
         $\Bar{U}_i'$ & Adjusted mean utilization for $\tau_i$ defined in~\cref{def: AAU}\\
         \hline
         $\Bar{U}_A'$ & Adjusted mean utilization for any task set $A$ defined in~\cref{def: AAU-taskset}\\
         \hline
         $\Phi$ & Deadline failure probability\\
         \hline
         $S_I$ & Sum of the execution times of jobs that are released and have their deadlines within an interval $I$\\
         \hline
    \end{tabular}
    
    \label{tab:symbol}
\end{table}

The assumed system consists of multiple homogeneous processing cores and DAG task set $\Gamma = \{\tau_1,...,\tau_n\}$. 
Each DAG task $\tau_i$ is defined by $(G_i,T_i,D_i, \rho_i)$, where $G_i = (V_i,E_i)$ is a DAG consisting of a set of nodes and a set of edges. 
$T_i$ and $D_i$ are positive integers and represent the period and relative deadline of the DAG task, respectively. 
The relationship between $T_i$ and $D_i$ is $D_i \leq T_i$, i.e., constrained deadline. 
$\tau_i$ is considered schedulable if the estimated deadline failure probability is below threshold~$\rho_i$~\cite{2023-FP-RT}. 
Similarly, a task set is considered schedulable if each estimated deadline failure probability in the task set is below the corresponding threshold.

\begin{figure}[t]
    \centering
    \includegraphics[trim= 30 30 50 50, clip, width=\linewidth]{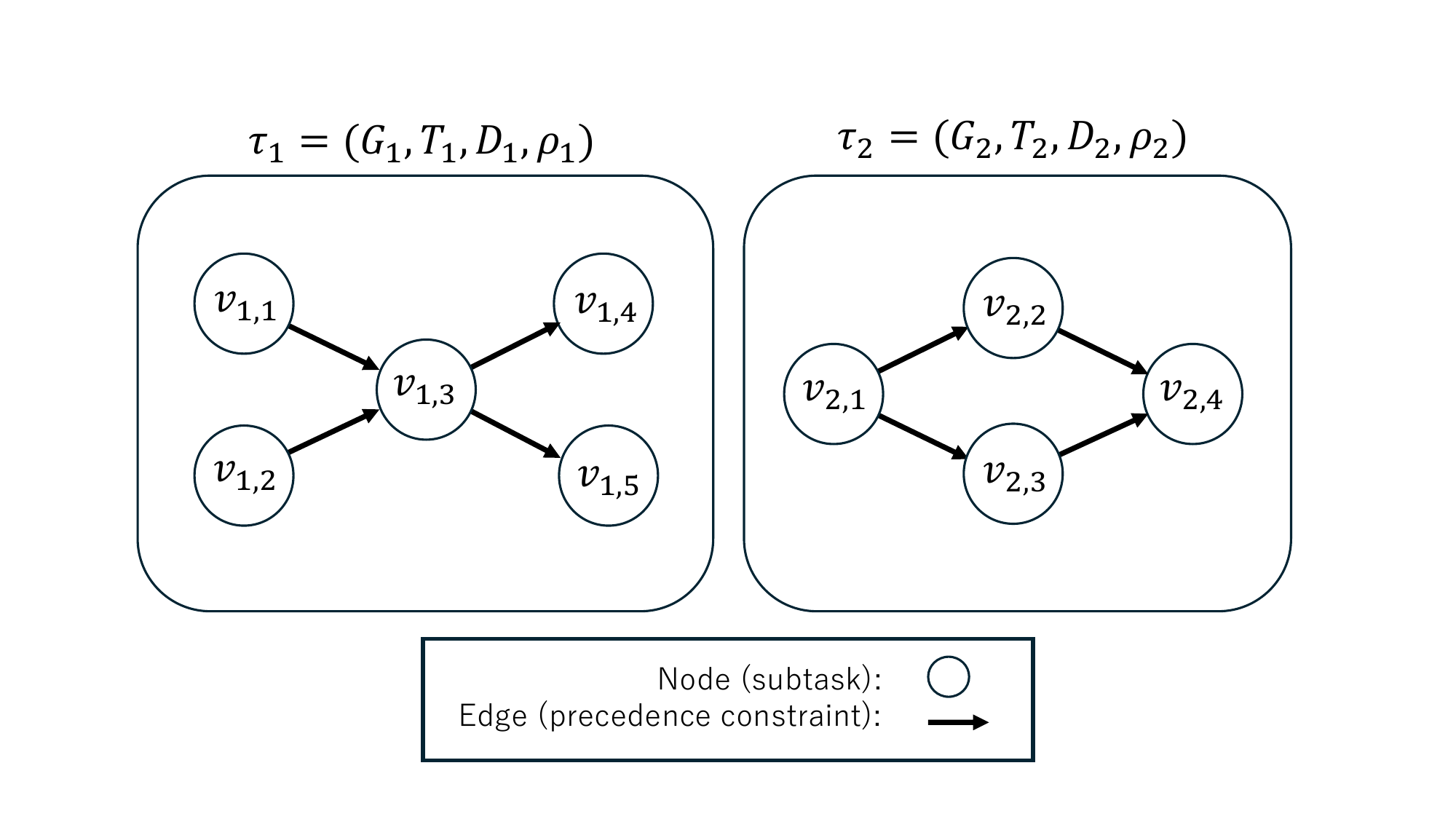}
    \caption{An example of the DAG task model, subtasks in the same frame have the identical relative deadline and the period.}
    \label{fig:taskModel}
\end{figure}

Each node $v_{i,j} \in V_i\quad( j\in \{1,...,|V_i|\})$ of the DAG is a subtask characterized by pWCET distribution $C_{i,j}$. 
The pWCET distribution $C_{i,j}$ is a discrete probability distribution where $C(x)$ is the probability of execution time $x$. A random variable with distribution $C$ is denoted as $\mathcal{C}$.
The period and relative deadline for each subtask of the DAG are inherited directly from the DAG task, and all jobs of each subtask are released sporadically and simultaneously with at least the period apart. 
Jobs past their deadlines are assumed to be aborted.
At any time, preemption is allowed. 
An edge $(v_{i,j},v_{i,k}) \in E_i$ signifies that the $\ell$-th job of subtask $v_{i,j}$ must complete before the $\ell$-th job of subtask $v_{i,k}$ starts.
An example of the system model is shown in~\cref{fig:taskModel}. 
Two DAG tasks are represented by the rounded boxes. 

Each DAG task is assigned to one processing core and scheduled under EDF. 
With this scheduling assumption, each DAG task $\tau_i$ can be represented as $(C_i, T_i, D_i, \rho_i)$, and $C_i$ is defined as follows: 

\begin{definition}[Probabilistic Total Execution Time Distribution]\label{def: PTETD}
    The probabilistic total execution time distribution $C_i$ of a task $\tau_i$ is defined using pWCET distribution of each node of the task and an operator $\bigotimes$ as
    \begin{equation}
        C_i = \bigotimes_{j=1}^{|V_i|}C_{i,j}.
    \end{equation}
    The operator $\bigotimes$ denotes convolution: for distribution $X$ and $Y$, $(X \otimes Y)(x) = \sum_{k=-\infty}^{\infty} \mathbb{P}(\mathcal{X}=k) \mathbb{P}(\mathcal{Y}=x-k) \quad for\quad x \in \mathbb{R}$.  
\end{definition}

The task $\tau_i$ can be represented as $(C_i, T_i, D_i, \rho_i)$ because sequentially scheduling nodes in a topological order of $G_i$ respects precedence constraints $E_i$.
Additionally, the probabilistic total execution time $C_i$ is independent of the execution order of the nodes in $V_i$ since the probabilistic execution time of each node $v_{i,j}\in V_i$ follows the independent execution time distribution $C_{i,j}$. 
%
The $j$-th job of $\tau_i$ is denoted as $J_{i,j}$, with absolute deadline $d_{i,j}$. 

Using probabilistic total execution time distributions (\cref{def: PTETD}), the following assumption is made for each task. 
\begin{assumption}\label{assumption}
    For each task $\tau_i$ included in $\Gamma$, the following inequality holds.
    \begin{equation}
        \mathbb{P}(\mathcal{C}_i > D_i) \leq \rho_i \label{eq: assumption}
    \end{equation}
    $\mathcal{C}_i$ is a random variable following $C_i$.
\end{assumption}
This assumption means that each task $\tau_i$ is schedulable if no other tasks are scheduled on the same processing core where $\tau_i$ is scheduled.

\section{Proposed Partitioning Algorithms}
\label{sec: proposed approch}
This section details the proposed approach: calculating the number of processing cores and partitioning a DAG task set $\Gamma$ under partitioned EDF. The section is organized as follows: utilization-related definitions for probability-aware partitioning (\cref{sec: prelim}), the proposed partitioning algorithm (\cref{sec: partition}), the schedulability analysis used (\cref{sec: analysis}), and the calculation of overload probabilities (\cref{sec: cal OP}).

\subsection{Preliminaries}\label{sec: prelim}
In this section, an adjusted mean utilization used in the proposed method is defined. 
To this end, an adjusted probabilistic total execution time distribution is defined as follows: 

\begin{definition}[Adjusted Probabilistic Total Execution Time Distribution]\label{def: APTETD}
    The adjusted probabilistic total execution time distribution $C_i'$ for a task $\tau_i$ is defined as
    \begin{equation}
        C_i'(c) = \mathbb{P}(\mathcal{C}_i'=c) = 
        \begin{cases}
            \mathbb{P}(\mathcal{C}_i=c) & \text{if} \quad c < D_i\\
            \sum_{k=c}^\infty \mathbb{P}(\mathcal{C}_i = k) & \text{if} \quad c = D_i\\
            0 & \text{if} \quad c > D_i
        \end{cases}
    \end{equation}
    $\forall c \in \text{Im } \mathcal{C}_i$ is an execution time.
\end{definition}

Using the adjusted probabilistic total execution time distribution, the adjusted mean utilization for a task and task set is defined as follows: 

\begin{definition}[Adjusted Mean Utilization of a Task]\label{def: AAU}
    Let $C_i'$ be the adjusted probabilistic total execution time distribution of $\tau_i$. The adjusted mean utilization $\Bar{U}_i'$ of $\tau_i$ is defined as
    \begin{equation}
        \Bar{U}_i' = \frac{\mathbb{E}[\mathcal{C}_i']}{T_i}.
    \end{equation}
\end{definition}

\begin{definition}[Adjusted Mean Utilization of a Task Set]\label{def: AAU-taskset}
    For any task set $A$, the adjusted mean utilization is defined as
    \begin{equation}
        \Bar{U}'_A = \sum_{\tau_i \in A} \Bar{U}_i'. 
    \end{equation}
\end{definition}
By~\cref{def: APTETD}, the adjusted mean utilization $\Bar{U}_i'$ for each task $\tau_i \in \Gamma$ is greater than zero and less or equal to one. 

The adjusted mean utilization accounts for the deadline-abort policy by disregarding deadline-exceeded jobs.
Unlike standard mean utilization, which can exceed one without \cref{def: APTETD} (e.g., $T_i = 5, C_i = \begin{pmatrix}1 & 10^3\\1-10^{-2} & 10^{-2}\end{pmatrix}\text{ then } U=\mathbb{E}[\mathcal{C}_i]/T_i=2.198 > 1$), the adjusted version remains within $(0,1]$, facilitating task set partitioning.

\subsection{Partitioning}\label{sec: partition}
This subsection details the proposed partitioning process, aiming to minimize processing cores needed to schedule $\Gamma$ while ensuring its schedulability.
For instance, if three tasks $\tau_1, \tau_2, \tau_3$ are individually schedulable (per \textbf{Assumption} in~\cref{sec: sys}), three cores are initially needed. If $\tau_1$ and $\tau_2$ can be scheduled together, the requirement reduces to two.

While exploring all partitioning patterns could minimize core count, the Bell number complexity limits this to small task sets . Thus, heuristics are necessary for practical, efficient partitioning with fewer divisions.

The proposed partitioning method uses ideas of the four heuristic methods, Item-Centric Best-Fit-Decreasing ($\mathbf{ICBFD}$), Item-Centric Worst-Fit-Decreasing ($\mathbf{ICWFD}$), Bin-Centric Best-Fit ($\mathbf{BCBF}$), Bin-Centric Worst-Fit ($\mathbf{BCWF}$). 
These four methods have been used as heuristics for the bin-packing problem~\cite{Baruah2013PartitionedEDF,bin-packing}. 
The prefixes IC (Item-Centric) and BC (Bin-Centric) denote two traditional bin-packing algorithm classes. IC considers each item to find a suitable bin based on a strategy (e.g., Best-Fit, Worst-Fit). BC considers each bin to find an item to fit according to a strategy for the current bin. 

A pseudo-algorithm for $\mathbf{ICBFD}$, which is Item-Centric, is shown in~\cref{algo: partition-IC,algo: find_core_id}.
\begin{algorithm}[t]
\SetKwInOut{Input}{Input}\SetKwInOut{Output}{Output}
    \Input{$\Gamma$}
    \Output{A partition of $\Gamma$} 
    $sorted\_partition \gets \{\{\}\}$; \tcp{List to contain lists of tasks}
    $sorted\_free\_tasks \gets sort\_by\_adjusted\_utilization(\Gamma)$; \tcp{Decreasing order} 
    \tcp{Head of sorted\_free\_tasks}
    \For{$task \in sorted\_free\_tasks$}
    {
        $core\_id \gets find\_core\_id(sorted\_partition, task)$\;
        \label{algo:line: find_core_id}
        \eIf{$core\_id$ is None}
        {
            add $\{task\}$ to $sorted\_partition$\;
        }
        {
            add $task$ to $sorted\_partition[core\_id]$\;
            \If{not $is\_feasible(sorted\_partition[core\_id])$\label{algo:line schedulable-ic}}
            {
                remove $task$ from $sorted\_partition[core\_id]$\;
                add $\{task\}$ to $sorted\_partition$\;
            }
        }
        \tcp{Decreasing order}
        $sorted\_partition \gets sort\_by\_adjusted\_utilization(sorted\_partition)$\;  
    }
    \Return $sorted\_partition$\;
    
\caption{Finding a partition with ICBFD}
\label{algo: partition-IC}
\end{algorithm}
\begin{algorithm}[t]
\SetKwInOut{Input}{Input}\SetKwInOut{Output}{Output}
    \Input{$sorted\_partition, task$}
    \Output{index of the core or $None$}
    $filtered\_core\_ids \gets \{i\in \{0,...,|sorted\_parition|-1\}:  \bar{U}'_{sorted\_partition[i]}+\bar{U}'_{task} \leq 1 \}$\;
    \eIf{$filtered\_core\_ids$ is $\emptyset$}
    {
        \Return $None$\;
    }
    {
        \tcp{Return the first element}
        \Return $filtered\_core\_ids[0]$\;
        \label{algo:line: return id}
    }
\caption{\textit{find\_core\_id} using Best-Fit}
\label{algo: find_core_id}
\end{algorithm}
In~\cref{algo:line: find_core_id} of~\cref{algo: partition-IC}, the function \textit{find\_core\_id} returns either the index of the core selected according to Best-Fit or \textit{None}. 
\textit{None} indicates that the index of the core cannot be selected according to Best-Fit.  
A pseudo-algorithm for \textit{find\_core\_id} in \cref{algo:line: find_core_id} of~\cref{algo: partition-IC} is shown in~\cref{algo: find_core_id}. 
\cref{algo: find_core_id} is an algorithm using Best-Fit-Decreasing, and the algorithm using Worst-Fit-Decreasing is one that returns the last element of \textit{filtered\_core\_ids} in~\cref{algo:line: return id} of~\cref{algo: find_core_id}.

A pseudo-algorithm for $\mathbf{BCBF}$, which is Bin-Centric, is shown in~\cref{algo: partition-BC,algo: best-fit}. 
\begin{algorithm}[t]
\newcommand{\prepare}[0] 
{
    add $cur\_subset$ to $partitioned\_tasks$\;
    $cur\_subset \gets \emptyset$\;
}
\SetKwInOut{Input}{Input}\SetKwInOut{Output}{Output}
    \Input{$\Gamma$}
    \Output{A partition of $\Gamma$}
    $cur\_subset$ $\leftarrow$ $\emptyset$; 
    $partitioned\_tasks$ $\leftarrow$ $\emptyset$\;

    $free\_tasks \leftarrow \Gamma$; \tcp{unassigned tasks}
    

    \While{$free\_tasks \neq \emptyset$}{
        \While{True}
        {
            
            $\tau \leftarrow find\_task(free\_tasks, cur\_subset$)\;
            \label{alg:line: find_task}
            \tcp{Task is not found}
            \If{$\tau = None$} 
            {
                \prepare
                continue\;
            }
            \tcp{Task is found}
            add $\tau$ to $cur\_subset$\;
            \tcp{All tasks in cur\_subset are schedulable}
            
            \If{$is\_feasible(cur\_subset)$\label{alg:line: schedulable}}
            {
                remove $\tau$ from $free\_tasks$\;
                break\;
            }

            \tcp{Not all tasks in cur\_subset are schedulable}
            remove $\tau$ from $cur\_subset$\;
            \prepare
            
        }
    }

    add $cur\_subset$ to $partitioned\_tasks$\;
    \Return $partitioned\_tasks$\;
    
\caption{Finding a partition with BCBF}
\label{algo: partition-BC}
\end{algorithm}

\begin{algorithm}[t]
\SetKwInOut{Input}{Input}\SetKwInOut{Output}{Output}
    \Input{$free\_tasks, cur\_subset$}
    \Output{one task or $None$}
    $filtered\_tasks \leftarrow \{\tau_i \in free\_tasks: \Bar{U}_{cur\_subset}'+\Bar{U}_i' \leq 1\}$\;
    \If{$filtered\_tasks = \emptyset$}
    {
        \Return $None$\;
    }
    \Return $\tau_i \leftarrow$ $\displaystyle\argmax_{\tau_i \in filtered\_tasks} \Bar{U}_i'$\;
    
\caption{\textit{find\_task} using Best-fit}
\label{algo: best-fit}
\end{algorithm}

In~\cref{alg:line: find_task} of~\cref{algo: partition-BC}, the function \textit{find\_task} returns either the task selected based on Best-Fit strategy or \textit{None}. 
\textit{None} indicates that the task could not be selected according to Best-Fit. 
A pseudo-algorithm for \textit{find\_task} in \cref{alg:line: find_task} of~\cref{algo: partition-BC} is shown in~\cref{algo: best-fit}. 
\cref{algo: best-fit} is an algorithm using Best-Fit. 
The algorithm that changed ``argmax'' to ``argmin'' in line~5 of~\cref{algo: best-fit} is one using Worst-Fit.

The function \textit{is\_feasible} in~\cref{algo:line schedulable-ic} of \cref{algo: partition-IC} and \cref{alg:line: schedulable} of \cref{algo: partition-BC} determines whether all the tasks given as the argument are schedulable in the sense described in~\cref{sec: sys}. 
Specifically, let $cur\_subset$ be a task set given by the argument and $DFP_i$ be the estimated deadline failure probability for $\tau_i \in cur\_subset$
The task set $cur\_subset$ is schedulable if $\forall \tau_i \in cur\_subset, DFP_i \leq \rho_i$. 
A specific method used in $is\_feasible$ is described in the next section.

\subsection{Schedulability Analysis}\label{sec: analysis}

This subsection details the schedulability analysis employed in the proposed approach, specifically within the $is\_feasible$ function (\cref{algo:line schedulable-ic} of~\cref{algo: partition-IC} and~\cref{alg:line: schedulable} of~\cref{algo: partition-BC}) to assess the schedulability of given tasks.

In the schedulability analysis, the existing method~\cite{EDF} is applied. 
This method calculates an upper bound of Worst-Case Deadline Failure Probability (WCDFP) under EDF. 
The definition of WCDFP is established in the existing research. 
WCDFP is based on the deadline failure probability (DFP) of a job. 
The DFP of a task $\tau_i$ is the largest probability among the DFPs of all jobs of $\tau_i$. 
Similarly, the DFP of a system is the largest probability among the DFPs of all the tasks. 
The WCDFP is the largest DFP for any job arrival sequence. 

In the existing method~\cite{EDF} of schedulability analysis used in the proposed approach, overload probabilities are calculated for specific intervals. 
The overload probability for an interval $I$ is the probability that the sum of execution times of jobs released and having deadlines within $I$ exceeds the interval length $|I|$. 
This can be expressed by the following formula. 
\begin{equation}
    \mathbb{P}(\mathcal{S}_I > |I|) 
\end{equation}
$I$ is the interval under consideration, and $\mathcal{S}_I$ is a random variable of the distribution $S_I$ obtained by convolving the execution time distributions of jobs released and having its deadlines within $I$. 
$S_I$ is calculated as follows:  
\begin{equation}
    S_I = \bigotimes_{\tau_i \in cur\_subset}^{N_i(I)} C_i
\end{equation}
$N_i(I)$ is the number of jobs released from $\tau_i$ and having its deadlines within $I$. 
\textit{cur\_subset} is the set of tasks to be analyzed given by the function \textit{is\_feasible} argument in~\cref{algo:line schedulable-ic} of \cref{algo: partition-IC} and \cref{alg:line: schedulable} of \cref{algo: partition-BC}. 

In the same research, a method was proposed to calculate the upper bound of the WCDFP of a task by extending periodic tasks to sporadic tasks and considering a worst-case arrival sequence that maximizes the DFP of a job. 
\begin{theorem}[Theorem 1 Ref.~\cite{EDF}]\label{th: WCDFP}
    For task $\tau_k$ in a given task set $part$, the task DFP $\Phi_k$ is bounded from above by
    \begin{equation}
        \Phi_k^* = \sum_{t_s \in A(0, H-D_k)} \mathbb{P}(S_{[t_s, H]}>H-t_s)\label{eq: WCDFP},
    \end{equation}
    where $A(0, H-D_k)$ is the set of all arrival times in $[0, H- D_k]$ when all tasks are aligned with the job $J_{k, H/T_k}$ with deadline $d_{k,H/T_k}$ = H. 
\end{theorem}
The worst-case arrival sequence that maximizes the DFP of $J_{i,j}$ is an arrival sequence where the deadline of one job for each task $\tau_k$ matches the deadline of $J_{i,j}$, and jobs prior to the jobs having the same deadline are released periodically with each minimal inter arrival time. 

As stated in the existing research, the WCDFP of a task obtained from~\cref{th: WCDFP} is the same value regardless of a specific task in $cur\_subset$ as the worst-case arrival sequence for each task is the same. 
Therefore, the WCDFP of a task obtained from~\cref{th: WCDFP} is the WCDFP of $cur\_subset$. 

In the existing research~\cite{EDF}, a method has been proposed to improve the analysis time by terminating the analysis midway while analyzing WCDFP using~\cref{th: WCDFP}. 
The idea of the method is described in Lemma~12 in Ref.~\cite{EDF}. 
\begin{lemma}[Lemma 12 Ref.~\cite{EDF}]\label{le: approx}
    For the arrival sequence $\alpha^*$ in \cref{th: WCDFP} and a given interval $[t_1, H]$ with $t_1 \in A(0, H - D_k)$, the portion $\Phi_k^{t_s <t_1}$ of the DFP $\Phi_k$ that is contributed by intervals $[t_s, H]$ with $t_s \in A(0, H - D_k)$ and $t_s < t_1$ is upper-bounded by
    \begin{equation}
        \Phi_k^{t_s < t_1} < \mathbb{P}_{busy}([t_1, H]),
    \end{equation}
    where $\mathbb{P}_{busy}([t_1, H])$ is the probability that the processor does not idle in the interval $[t_1,H]$. 
\end{lemma}
A summary of the method to improve the analysis time is to stop the calculation when the probability $P_{busy}(I)$ that the currently considered interval $I = [t_x, H]$ will be busy sinks below a given threshold value. 
Specifically, let $\Phi_i^+$ denote the current estimate of $\Phi_i^*$ when encountering $t_x$ in the computation of \cref{eq: WCDFP}. 
Once the threshold is exceeded, $\Phi_i^*$ is set to $\Phi_i^+ + \mathbb{P}_{busy}(I)$.
Since the method terminating the analysis midway has been shown to significantly improve analysis time in the evaluation of the research~\cite{EDF}, this method is also used in the schedulability analysis in~\cref{algo:line schedulable-ic} of \cref{algo: partition-IC} and \cref{alg:line: schedulable} of \cref{algo: partition-BC}.

\subsection{Our Approach to Calculating Overload Probability}\label{sec: cal OP}
In our proposed method, $S_I$ explained in~\cref{sec: analysis} is obtained using a method not used in the existing study~\cite{EDF}, and the overload probability is finally calculated.
\cref{th: WCDFP} and \cref{le: approx} allow us to calculate the WCDFP. 
To allow the analysis to be truncated midway using \cref{le: approx}, the overload probability for the interval $[t_s, H]$ is calculated in descending order of $t_s$ in \cref{th: WCDFP}. 
Let $t_s'$ be the next arrival time to consider the overload probability after calculating the overload probability for $[t_s, H]$. 
Considering the jobs in the interval $[t_s, H]$ and the jobs in the interval $[t_s', H]$, the only difference is the jobs released at $t_s'$. 
Therefore, when calculating $S_{[t_s', H]}$, $S_{[t_s, H]}$ is utilized as follows: 
\begin{equation}\label{eq: iter}
    S_{[t_s',H]}= S_{[t_s, H]} \otimes \bigotimes_{\tau_i \in Tasks(t'_s)} C_i
\end{equation}
$Tasks(t_s')$ is the set of all the tasks that release jobs at $t_s'$. 
By using~\cref{eq: iter}, it is no longer necessary to convolve the execution time distributions of all jobs in the interval $[t_s',H]$ when calculating $S_{[t_s',H]}$.

\section{Application of the Proposed Method to Federated Scheduling Approach}
\label{sec: app2fed}
In this section, a method to utilize the proposed method in~\cref{sec: proposed approch} for federated scheduling is proposed. 
The advantage of this method is that the proposed method can improve the federated scheduling method with high schedulablity by improving the conventional excessive core requirements because the proposed method can perform probabilistic analysis.
Federated scheduling~\cite{first-fed} is an approach to efficiently schedule parallel real-time tasks on a multicore processor.
In federated scheduling, the task set is classified into heavy tasks and light tasks, and each heavy task is assigned an exclusive core for analysis, while light tasks are analyzed using cores left over from the analysis of heavy tasks.
The flow of the proposed method applied to federated scheduling is as follows.
\begin{enumerate}
    \item Classify DAG tasks satisfying \cref{eq: assumption} as light; the others as heavy.
    \item Calculate required cores $m_{heavy}$ for heavy tasks using any suitable analysis.
    \item Calculate required cores $m_{light}$ for light tasks using our proposed method from~\cref{sec: proposed approch}.
    \item Declare the task set schedulable if $m_{heavy} + m_{light}$ is within the available core count.
\end{enumerate}

\section{Evaluation}
\label{sec: evaluation}
The performance of the proposed task set partitioning method is experimentally evaluated in terms of schedulability under varying total DAG task set utilization. This section details the compared existing method (\cref{sec: compared method}), the DAG task set generation setup (\cref{sec: setup}), comparisons with non-probabilistic partitioning (\cref{sec: cmp to wc}), comparisons with the existing method detailed in~\cref{sec: compared method} (\cref{sec: cmp w existing method}), comparisons among different heuristics (\cref{sec: cmp heu}), a comparison of federated scheduling (\cref{subsec: cmp-fed}), and a summary of findings (\cref{sec: LL}).

\subsection{Compared Method}\label{sec: compared method}
Among the existing schedulability analysis methods for DAG tasks with probabilistic execution times~\cite{DAGprob,SHPC-DAG,DAG-E2E}, this section focuses on the most comparable one.
The compared existing method~\cite{DAGprob} calculates an upper bound on the probabilistic response time of a DAG task through three steps: local response time (LRT), response time in isolation (RTI), and global response time (GRT). These times are iteratively calculated for each node from source to sink, considering predecessor nodes (LRT), parallel nodes within the same DAG (RTI), and higher-priority DAG tasks (GRT). The GRT of the sink node represents the response time of the DAG task.

This method~\cite{DAGprob} requires a core and priority assignment for each DAG node but does not propose a core assignment method.
In the evaluation of this paper, to apply this existing method to the function $is\_feasible$ in~\cref{algo:line schedulable-ic} of \cref{algo: partition-IC} and \cref{alg:line: schedulable} of \cref{algo: partition-BC}, the core allocation is determined by~\cref{algo: partition-IC}, and nodes of the same DAG task are assigned to the same core.
Comparison with the proposed method will be conducted in~\cref{sec: cmp w existing method}.

\subsection{Evaluation Setup}\label{sec: setup}
The evaluation uses randomly generated DAG task sets, with individual DAG tasks generated using “fan-in/fan-out” feature of RD-Gen~\cite{RD-Gen} adapted for our model.
The number of source and sink nodes is fixed to one, and the in-degree and out-degree of the DAG are set to one through three at random.
The period for each task is from 1,000 to 10,000 in increments of 1,000. 
The number of nodes in each DAG task takes the value ranging from two to five. 
The worst-case utilization of each DAG task is from 0.5 to 1.0 in increments of 0.1 in~\cref{sec: cmp w existing method} and 0.1 to 1.0 in~\cref{sec: cmp heu,sec: cmp to wc}. 
The task set is generated by randomly selecting tasks until reaching the target task set utilization.

To fit the generated DAGs to our model, the probabilistic execution time for each node is set based on the given WCET. 
The probability of the possible execution time is set to $\mathbb{P}(WCET/3)=0.98$ and $\mathbb{P}(WCET)=0.02$, reflecting the empirical distribution of Autoware runtimes observed in CARET measurements, as utilized in the evaluation settings of Ref.~\cite{ECRTS-toba}.
The parameters for the evaluation are listed in~\cref{tab: parameter}. 
\begin{table}[t]
    \centering
    \caption{Task set generation parameters and evaluation system configurations}
    \begin{tabular}{|l|p{3.6cm}|}
        \hline
        \textbf{Task Parameter} & \textbf{Description}\\ 
        \hline
         Period & value from 1,000 to 10,000 in increments of 1,000\\
         \hline
         Num. of nodes & value from two to five\\
         \hline
         Num. of source nodes and sink nodes & one\\
         \hline
         In-degree and out-degree & value from two to five\\
         \hline
         Worst-Case Utilization & value from 0.5 to 1.0 in increments of 0.1 in~\cref{sec: cmp w existing method} and 0.1 to 1.0 in~\cref{sec: cmp heu,sec: cmp to wc}\\
         \hline
         Threshold $\rho$ & 0.0001\\

         \hline\hline
         \textbf{Task Set Parameter}& \textbf{Description}\\
         \hline
         Worst-Case Utilization &  value from 0.8 to 8.0 in increments of 0.4\\
         \hline\hline
         \textbf{Other Parameter} & \textbf{Description}\\
         \hline
         Num. of cores & four\\
         \hline
         Num. of task sets at each utilization & 1,000\\
         \hline
         Analysis timeout duration per task set & 1,000 s\\
         \hline
    \end{tabular}
    
    \label{tab: parameter}
\end{table}
The threshold $\rho$ of each DAG task is set to 0.0001. 
The system is assumed to have four cores. 
To perform the evaluation on many task sets and improve the reliability of the evaluation, the timeout duration of the analysis per task set is set to 1,000 s.

To perform the evaluation, all the methods are written in Python and executed on a system with a 5.3 GHz 24-Core AMD Ryzen Threadripper 7960X CPU, 256 GiB RAM, and Ubuntu 22.04 LTS (64-bit) OS.

\subsection{Comparison with WCET-based Method}\label{sec: cmp to wc}
\begin{figure}[t]
    \centering
    \includegraphics[width=\linewidth]{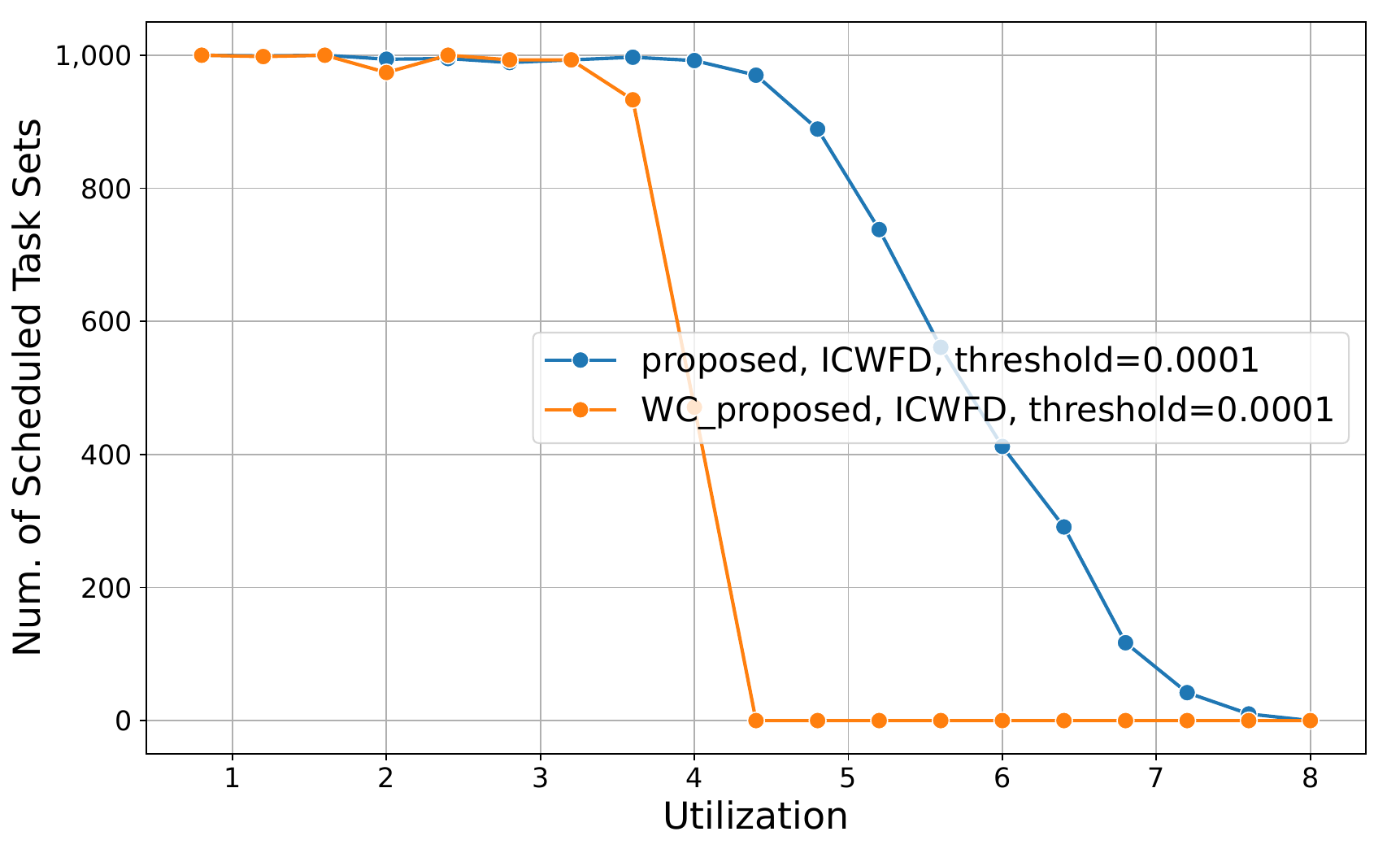}
    \caption{Comparison of the number of scheduled task sets between the proposed method and the method not considering probabilistic execution times.}
    \label{fig: comparison-wc-utili}
\end{figure}

A comparison of the number of scheduled task sets with $\mathbf{ICWFD}$ without considering probabilistic execution times during partitioning is shown in~\cref{fig: comparison-wc-utili}. 
In the case of ``WC\_proposed,'' which is based on worst-case execution times and does not consider  probabilistic execution times,
the number of scheduled task sets is zero if the task set utilization is greater than four, because task sets that exceed the utilization of four do not fit on four cores no matter how the task sets are configured.  
When the utilization is less than four, ``proposed'', which considers probabilistic execution times, is able to schedule almost all task sets, but the method that considers only worst-case execution times is unable to schedule some task sets at the utilization of 3.6 and 4.0. 
This is considered to be because $\mathbf{ICWFD}$ without considering probabilistic execution times judges that even if a core still has room, this method sometimes does not allow selection of tasks to be assigned to the core. 
The reason why the number of scheduled task sets at utilization of 2.0 is not 1,000 is because some task sets were determined to be unschedulable due to timeouts. 

\subsection{Comparison with Existing Method}\label{sec: cmp w existing method}

\begin{figure}[t]
    \centering
    \includegraphics[width=\linewidth]{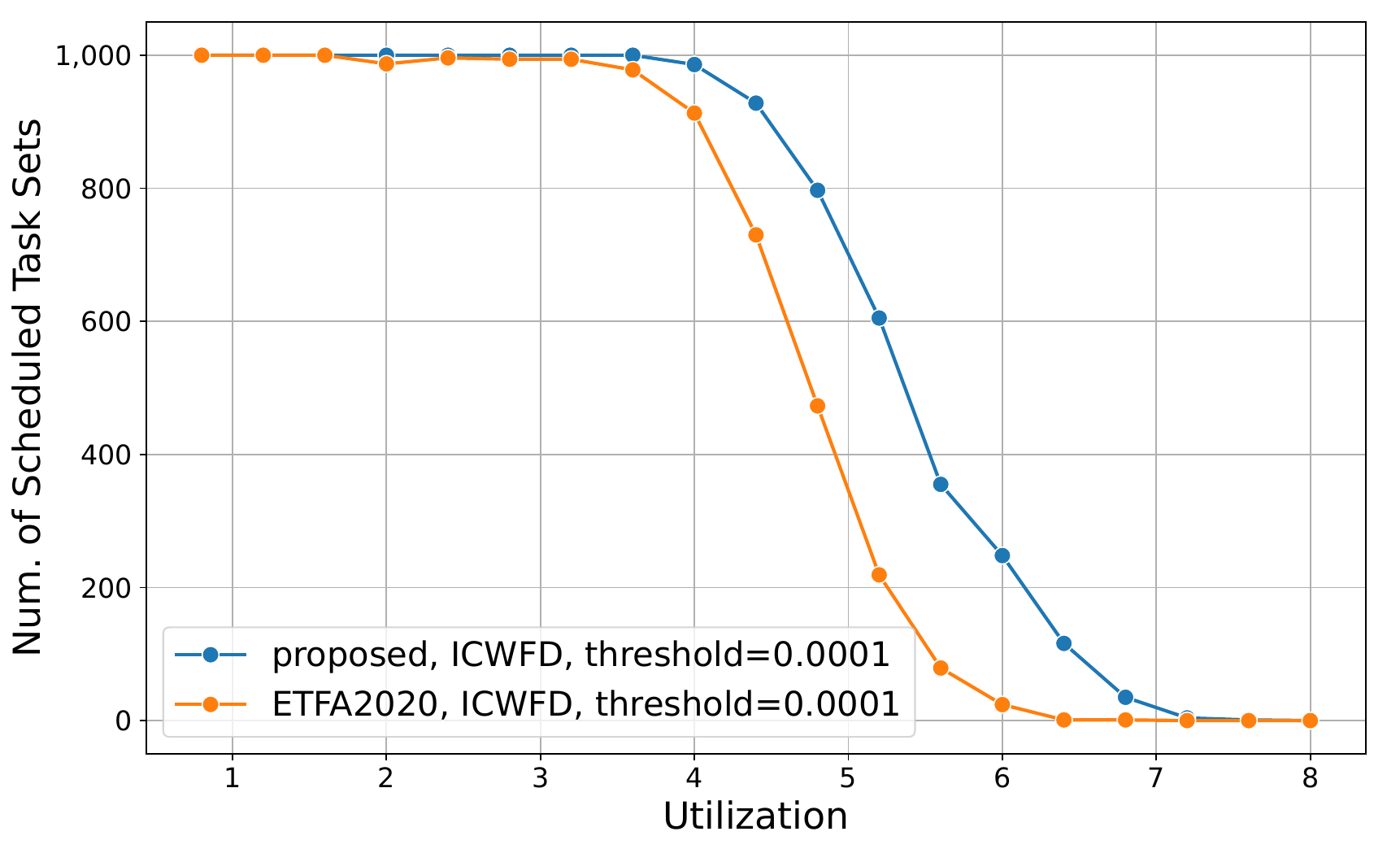}
    \caption{Comparison of the number of scheduled task sets between the proposed method and the existing method.}
    \label{fig: comparison-utili}
\end{figure}

\begin{figure}[t]
    \centering
    \includegraphics[width=0.8\linewidth, trim=0cm 0cm 0cm 0cm, clip]{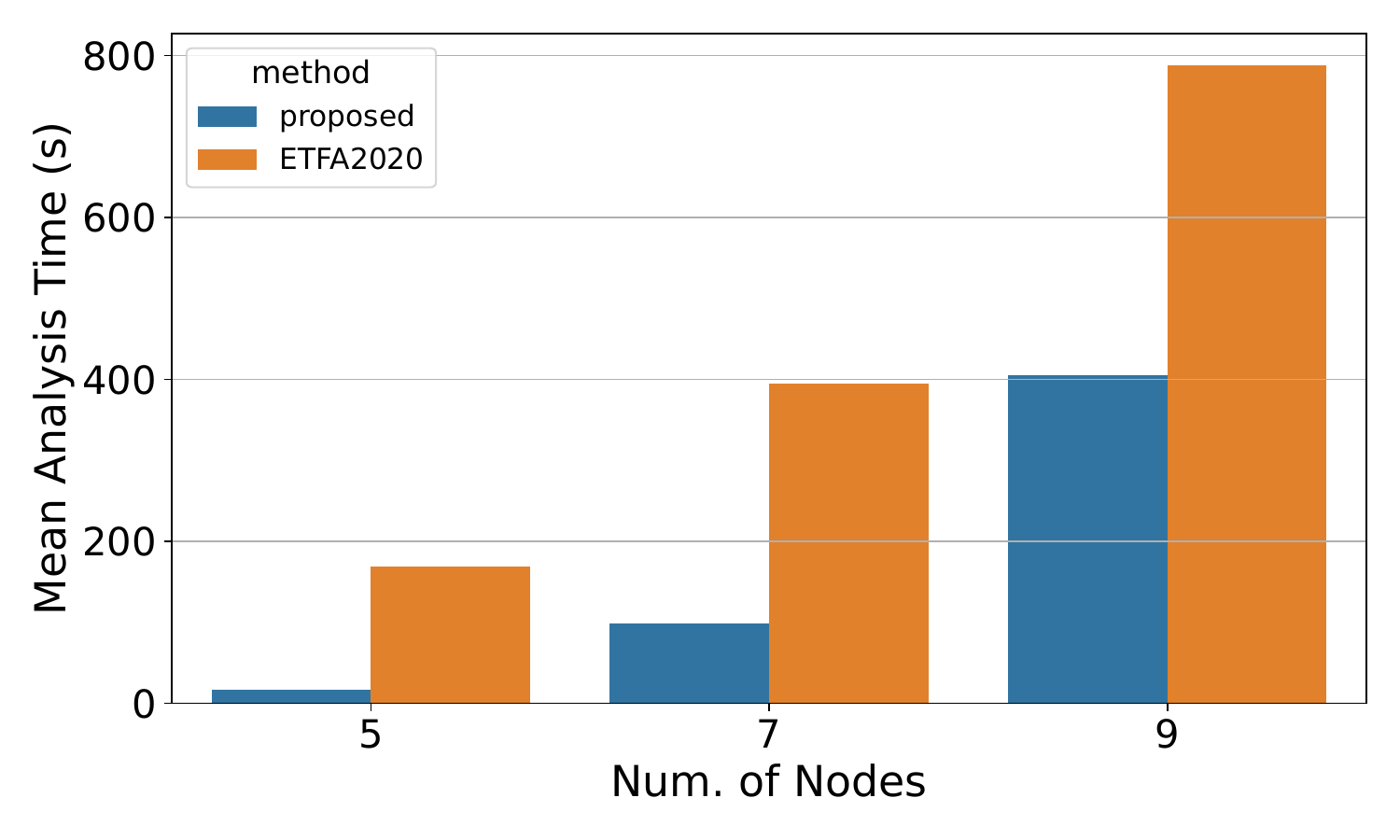}
    \caption{Comparison of the mean analysis time between the proposed method and the existing method for varying the number of nodes in each DAG.}
    \label{fig: comparison-time-mean}
\end{figure}

The results of the evaluation with the existing method described in~\cref{sec: compared method} are shown in~\cref{fig: comparison-utili,fig: comparison-time-mean}. 
``proposed'' in the legend  is the proposed method described in~\cref{sec: proposed approch}, and ``ETFA2020'' is the existing method explained in~\cref{sec: compared method}. 
The number of scheduled task sets is compared in \cref{fig: comparison-utili} for the case where the partitioning method is $\mathbf{ICWFD}$. 
As the analysis time increases with the number of tasks in the task set, the possible utilization for each task is limited to a range of 0.5 to 1.0.
\cref{fig: comparison-utili} shows that the proposed method ``proposed'' can scheduled the same or a larger number of task sets than the existing method ``ETFA2020''. 
In particular, at the utilization of 5.2, the proposed method schedules 38.6\% (386 task sets) more task sets than the existing method.
Task sets with analysis times exceeding 1,000 s were considered unschedulable although the percentage of task sets actually considered unschedulable was at most 1.3\% (13 task sets) in ``ETFA2020'' with the utilization of 2.0.

\begin{figure}[t]
    \centering
    \includegraphics[width=0.9\linewidth, trim=0cm 0cm 0cm 0cm, clip]{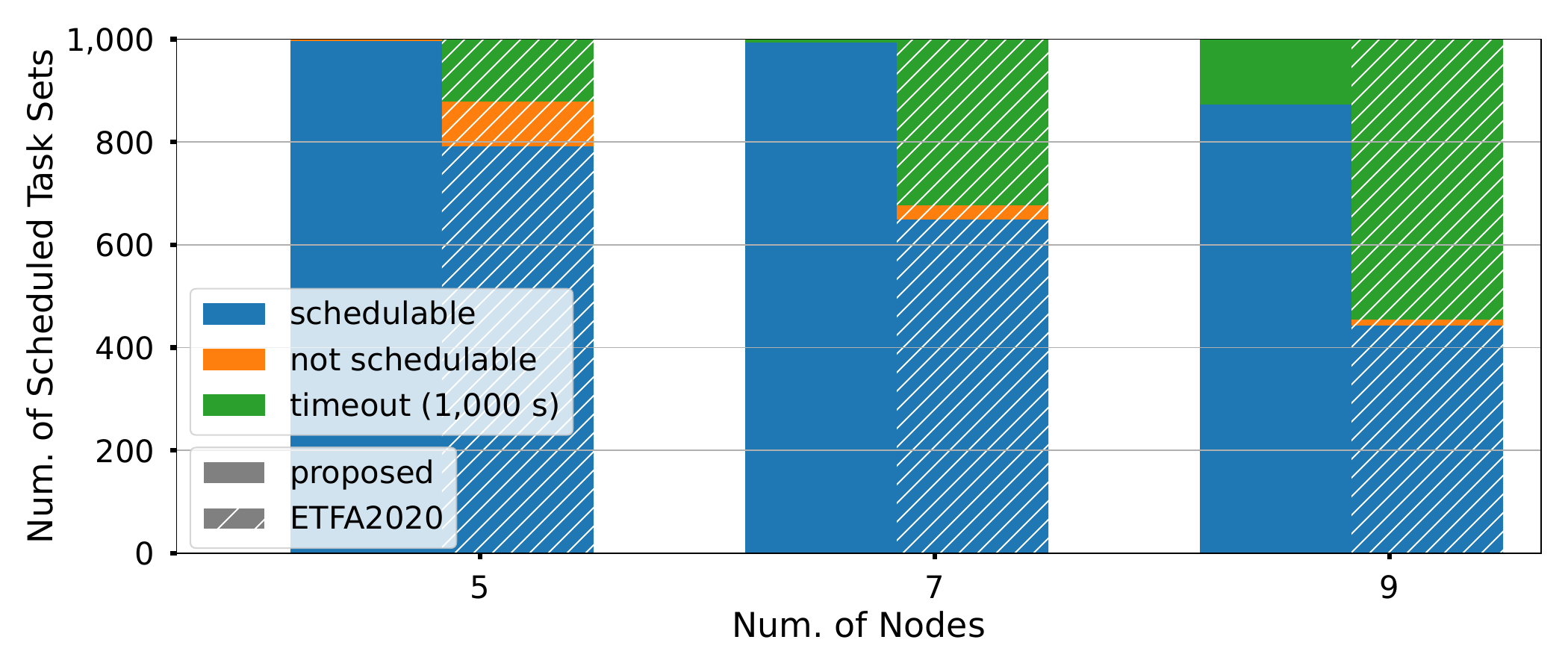}
    \caption{Comparison of task set classification between the proposed method and the existing method.}
    \label{fig: comparison-class}
\end{figure}

A comparison of the mean analysis time per task set for varying numbers of nodes in each DAG is shown in~\cref{fig: comparison-time-mean}, and the same task sets, classified by schedulability, is shown in~\cref{fig: comparison-class}. 
Although the analysis time for each task set that timed out is more than 1,000~s, the analysis time is treated as 1,000~s for the mean calculation.
The proposed method is shown to have shorter mean analysis times on the task set for varying number of nodes in each DAG. 
The ratio of the mean analysis time of ``proposed'' to the mean analysis time of ``ETFA2020'' becomes smaller as the number of nodes increases, and the difference in the mean analysis time seems to be smaller.
However, looking at the classification of the task sets used in the evaluation in~\cref{fig: comparison-class}, the percentage of task sets judged to be timed out for ``ETFA2020'' is more than 25\% for node number seven and more than 50\% for node number nine.
This means that the mean analysis time without timeout is larger than the mean analysis time with timeout, which was the configuration used for the evaluations in this paper.

\subsection{Comparison of Different Heuristics for Proposed Method}\label{sec: cmp heu}

\begin{figure}[t]
    \centering
    \includegraphics[width=\linewidth]{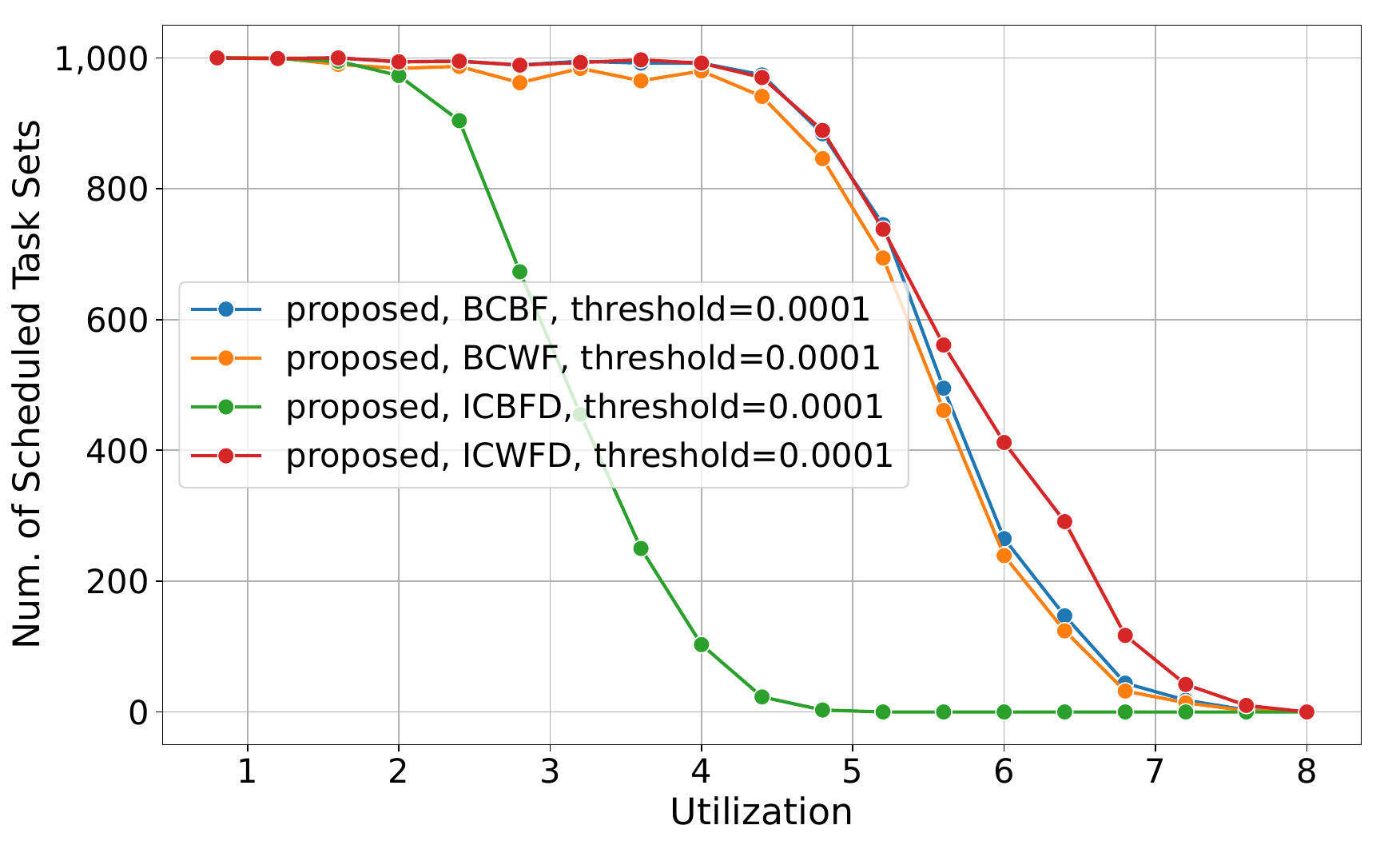}
    \caption{Comparison of the number of scheduled task sets among different heuristics.}
    \label{fig: comparison-algo-utili}
\end{figure}

The four heuristics $\mathbf{ICBFD}$, $\mathbf{ICWFD}$, $\mathbf{BCBF}$, and $\mathbf{BCWF}$ described in~\cref{sec: partition} are evaluated in terms of the number of scheduled task sets is shown in~\cref{fig: comparison-algo-utili}. 
The number of scheduled task sets for the three methods other than $\mathbf{ICBFD}$ are relatively close to each other as the task set utilization changes. 
The reason $\mathbf{ICBFD}$ takes a lower value than the other methods is thought to be because the method assigns tasks to a core that has no spare capacity. 
$\mathbf{ICBFD}$ allocates tasks to the core with the highest utilization not exceeding one.
As a result, the task set on the core tends to become unschedulable in the schedulability analysis.
Furthermore, in the Item-Centric algorithm, cores that have been determined to be unallocatable in the past and cannot afford to be allocated are also included in the core search range of the Item-Centric algorithm.
%
For example, with cores at utilizations of 0.7 and 0.3, ICBFD would choose the 0.7 core for a 0.2 utilization task, even though the 0.7 core is barely schedulable.
When considering the allocation of a task with a utilization of 0.2, $\mathbf{ICBFD}$ chooses the core with the utilization of 0.7, which cannot afford to schedule an additional task. 
Since the tasks to be allocated are selected in such a way that the utilization decreases monotonically, the selection of cores with no room to spare is repeated, resulting in a lower value for the number of scheduled task sets than in the other three methods. 

The three methods schedule more task sets in the order of $\mathbf{ICWFD}$, $\mathbf{BCBF}$, and $\mathbf{BCWF}$. 
Since $\mathbf{ICWFD}$ explores all active cores and allocates a task so that they have the smallest utilization when the task is assigned, this strategy seems to make sense in terms of preserving schedulability. 
$\mathbf{BCBF}$ and $\mathbf{BCWF}$ are Bin-Centric algorithms, which assign as many tasks as they can to the current core based on the strategy, then move to the next core and repeat with the same process. 
The difference between the two methods is less than 5\%.

\subsection{Comparison of Existing Federated Scheduling and Our Federated Scheduling}\label{subsec: cmp-fed}
In this subsection, our federated scheduling proposed in~\cref{sec: app2fed} is compared with an existing and state-of-the-art federated scheduling method~\cite{SOA-fed}. The schedulability analysis of heavy tasks in our federated scheduling proposed in~\cref{sec: app2fed} uses the heavy tasks analysis method of Ref.~\cite{SOA-fed}.

The existing research~\cite{SOA-fed} clearly defined DAG parallelism to derive a response time upper bound for DAG tasks and proposed its application to federated scheduling. This method showed better schedulability by utilizing DAG chain decomposition, considering node execution times.
This method~\cite{SOA-fed} classifies DAG tasks with $ vol(G_i)> D_i$ as heavy and others as light, calculating heavy task core requirements per Algorithm~2 in Ref.~\cite{SOA-fed}. $vol(G_i)$ is the sum of node WCETs. Light tasks are scheduled using the existing partitioned EDF~\cite{light-task-fed}.

The evaluation assumes an eight-core environment and per-DAG task utilization from 0.1 to 2.0 (0.1 increments) for heavy task generation. Task sets had a number of random nodes ranging from five to nine, with worst-case utilizations from 1.0 to 12.0 across 500 DAG tasks. The other settings match~\cref{tab: parameter}.

\begin{figure*}[t]
    \centering
    \includegraphics[width=\linewidth]{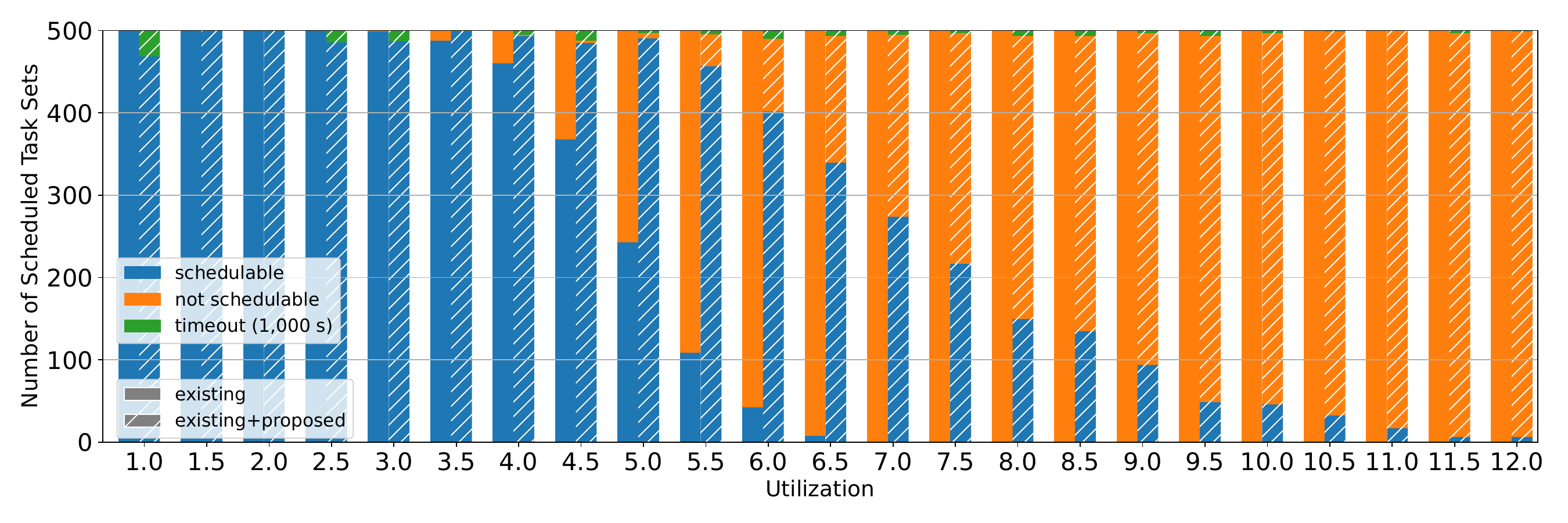}
    \caption{Comparison of task set classification between our federated scheduling and the state of the art federated scheduling.}
    \label{fig: comparison-util-class-fed}
\end{figure*}
The evaluation result of the proposed method in~\cref{sec: app2fed} and the latest federated method~\cite{SOA-fed} are shown in~\cref{fig: comparison-util-class-fed}. The horizontal axis shows the utilization of task sets, and the vertical axis shows the number of task sets. The label ``existing" is the latest federated method~\cite{SOA-fed} and ``existing+proposed" is the proposed method in~\cref{sec: app2fed}. The ratio of scheduled task sets in~\cref{fig: comparison-util-class-fed} shows that the proposed method is able to schedule more task sets than the existing method from a task utilization of 3.5 or higher. Since an eight-core environment is assumed, any task set that exceeds a task set utilization of 8.0 on the horizontal axis cannot be scheduled by the existing method that does not consider probabilistic execution time. On the other hand, the proposed method can schedule several task sets with a utilization that exceeds the number of cores because this method considers probabilistic execution time. 
At utilizations of 3.0 and 3.5, several task sets are judged to be timed out by the proposed method. This occurs because the schedulability analysis of the light tasks, i.e., the proposed method in~\cref{sec: analysis}, is time consuming.

\subsection{Lessons Learned}\label{sec: LL}
The effectiveness of adjusted mean utilization was confirmed through comparisons with worst-case utilization.
By using a method that utilizes the adjusted mean utilization that takes into account the probability of task execution time occurrence, \cref{fig: comparison-wc-utili} demonstrated that scheduling can be performed even when the worst-case utilization of a task set exceeds the number of cores used for scheduling. 

The higher schedulability of the proposed method and its applicability to the analysis of more complex task sets in terms of the mean analysis time were confirmed by comparing the proposed method with the existing schedulability analysis method for DAG task sets.
Comparing the schedulability performance of the proposed and the existing method, \cref{fig: comparison-utili} showed that the proposed method can schedule a larger number of task sets. 
Increasing the number of nodes in each DAG task, the mean analysis time was still reduced compared to the existing method, and \cref{fig: comparison-time-mean,fig: comparison-class} showed that the proposed method can analyze even such a complex task set.
It should be noted that in Fig. 4 any task set that times out at 1,000 s is counted as having an analysis time of exactly 1,000 s, which may bias the reported average analysis time downward.

$\mathbf{ICWFD}$ was identified as the best heuristics to use in the proposed method in terms of the schedulability, based on the comparison of four heuristics: $\mathbf{ICBFD}$, $\mathbf{ICWFD}$, $\mathbf{BCBF}$, and $\mathbf{BCWF}$.
Comparing the schedulability of the four methods, \cref{fig: comparison-algo-utili} shows that $\mathbf{ICWFD}$, $\mathbf{BCBF}$, $\mathbf{BCWF}$, and $\mathbf{ICBFD}$ can schedule more task sets, in that order.
As a conclusion on which method should be used, 
$\mathbf{ICWFD}$ is the best method in terms of the percentage of scheduled task sets 
compared to the remaining three methods. 

By using the proposed method, the performance of the schedulability analysis of DAG tasks that cannot be scheduled on one core was confirmed to be improved. 
The applicability of the proposed method in~\cref{sec: proposed approch} is restricted to a certain class of DAG tasks, subject to \textbf{Assumption} made in~\cref{assumption}.
However, by utilizing the proposed method in federated scheduling as proposed in~\cref{sec: app2fed}, tasks that cannot be handled due to ~\textbf{Assumption} can be handled, and \cref{fig: comparison-util-class-fed} shows that the proposed method can schedule more task sets than the existing federated scheduling method~\cite{SOA-fed}.



\section{Related Work}
\label{sec: related-work}
\begin{table}[t]
    \centering
    \caption{Comparison of the proposed method with related work}
    \begin{tabular}{|l|c|c|c|c|c|}
         \hline
         \textbf{Method}& $\textbf{EDF}^{a)}$ & $\textbf{SA}^{b)}$  & $\textbf{DAG}^{c)}$ & $\textbf{MC}^{d)}$ & $\textbf{NDC}^{e)}$\\
         \hline
          RTSS 2021~\cite{EDF}& \checkmark &  \checkmark  &&&\\
          \hline
          IEEE TC 2022~\cite{DAG-E2E} & \checkmark & \checkmark & \checkmark & \checkmark&\\
          \hline
          JSA 2024~\cite{SHPC-DAG}& \checkmark & \checkmark & \checkmark & \checkmark &\\
          \hline
          ETFA 2020~\cite{DAGprob}&  & \checkmark &\checkmark &\checkmark & \checkmark\\
          \hline
          RTSS 2023~\cite{srv-dag}& \checkmark&  & \checkmark & \checkmark & \checkmark \\
          \hline
          Proposed method & \checkmark & \checkmark &\checkmark &\checkmark & \checkmark\\
         \hline
    \end{tabular}
    \begin{flushleft}
        a) EDF: Adopt EDF policy within the system model\\
        b) SA: Schedulability analysis can be performed\\
        c) DAG: DAG model is used as the task model\\
        d) MC: A multi-core environment is assumed as the system model\\
        e) NDC: Nodes of different DAGs can be scheduled on the same core 
    \end{flushleft} 
    \label{tab: contribution}
\end{table}

This section reviews related work on task partitioning, probabilistic schedulability analysis, and federated scheduling.

\subsection{Partitioning}\label{sec: RW-partition}

Multicore scheduling~\cite{dag-survey} includes global scheduling (task migration allowed) and partitioned scheduling (tasks assigned to specific cores). Partitioning typically uses heuristic-based bin-packing for core assignment. 
Research in this area includes an energy-efficient method considering CPU frequency and utilization~\cite{energy-efficient-partition}, and an approach~\cite{harmonic-partitioning} leveraging task periods for Rate Monotonic scheduling.

Research~\cite{Baruah2013PartitionedEDF} proposed a speedup factor-based metric for evaluating core assignment heuristics, arguing its superiority over utilization-bound metrics. This metric indicates that effective partitioning aims for high processor utilization with a low speedup factor, approaching optimality. Reasonable Allocation Decreasing (RAD) algorithms, sorting tasks by non-increasing utilization, are considered optimal by this metric and include Worst-Fit-Decreasing, employed in our proposed method.

\subsection{Probabilistic Schedulability Analysis}\label{sec: RW-SA}
Probabilistic schedulability analysis, handling multiple states due to probabilistic execution times, can be intractable for large task sets. Resampling and statistical analysis~\cite{probability-survey} mitigate this.
Resampling reduces state space by decreasing the number of execution time-probability pairs. Statistical analysis estimates the probability of deadline exceeding based on statistical features such as means and variances of task execution times, avoiding direct convolution.
To reduce the computational complexity of convolution, circular convolution using the Fourier transform is proposed in Ref.~\cite{circle-convo}. 
Traditional linear convolution has a computational complexity of $O(n^2)$, which becomes a bottleneck when dealing with large-scale probability distributions. 
In contrast, circular convolution uses the Fourier Transform to perform convolution efficiently with a complexity of $O(n \log n)$.

For fixed priority scheduling on a single processor, response time analysis utilizes the critical instant concept~\cite{critical}. However, the research indicates that the direct extension of the deterministic critical instant theorem to the probabilistic domain does not necessarily represent the worst-case scenario. This research also identifies critical instants applicable to probabilistic execution times.

Under the EDF scheduling on a single processor, a schedulability analysis method~\cite{EDF}, which is incorporated into the proposed method, has been proposed. 
This analysis guarantees schedulability by considering the job arrival sequence in which a job is most probably to make a deadline miss and calculating the deadline failure probability for that job.

In contrast to the extensive deterministic schedulability analysis for DAG tasks~\cite{dag-survey}, probabilistic analysis for DAG tasks is limited. Beyond schedulability analysis, several research focuses on DAG task models with probabilistic execution times characterized nodes~\cite{srv-dag,srv-dag-ext}.
Ref.~\cite{srv-dag} proposed a server budget management method across the DAG and a method to analyze the probability of DAG task abortions under server-based scheduling, aiming to reduce WCET overruns and abort rates. This budget management improves the abortion probability. Ref.~\cite{srv-dag-ext} refined the budget management policy from Ref.~\cite{srv-dag} by limiting the parallel execution assumption of successive DAG instances.

To the best of our knowledge, research on schedulability analysis in DAG tasks with probabilistic execution time is three studies~\cite{DAG-E2E,SHPC-DAG,DAGprob}. 
The research~\cite{DAG-E2E} analyzes end-to-end latency in autonomous driving systems by modeling them as a DAG. The proposed method~\cite{DAG-E2E} divides the DAG into subgraphs based on node periodicity, analyzes subgraph response time distributions, and derives the overall latency distribution. Since this method requires at least one core to be assigned to each subgraph, nodes in different subgraphs cannot share cores, making it incomparable to our proposed method in \cref{sec: evaluation}.
%
The research~\cite{SHPC-DAG} proposes a scheduling approach using the SHPC-DAG model and LLP algorithm for real-time applications with uncertainties, optimizing a Safety-Performance metric under probabilistic execution times and conditional nodes. 
%
The Research~\cite{DAGprob}, compared with our method in~\cref{sec: cmp w existing method}, analyzes DAG tasks with probabilistic execution times, proposing a response time analysis assuming that fixed priority and core assignment for each DAG node are given, deriving response times through the convolution of execution time distributions using an iterative equation.

\subsection{Federated Scheduling}
Federated scheduling is an approach for parallel real-time tasks.
Recent research covers mixed-criticality~\cite{mixed-cri-fed}, arbitrary-deadline DAG~\cite{new-fed}, and conditional DAG~\cite{SOA-fed} federated scheduling.
For probabilistic scenarios, a federated scheduling method~\cite{pro-fed} exists for DAG tasks with probabilistic execution time.
This method classifies tasks as heavy or light based on the expected probabilistic execution time, guaranteeing a bounded expected delay, which quantifies deadline exceedance of the tasks.

\section{Conclusion}
\label{sec: concl}
This paper proposed a partitioning method for guaranteeing schedulability under partitioned EDF for pWCET-based DAG tasks. The method partitions tasks based on adjusted mean utilization and performs schedulability analysis during core assignment. By treating DAG tasks as single-node tasks, the state-of-the-art EDF schedulability analysis~\cite{EDF} for single processors is applicable. 
The effectiveness of adjusted mean utilization was confirmed by comparison with a WCET-based partitioning method.
The experimental evaluation showed that the proposed method outperforms the existing DAG schedulability analysis method~\cite{DAGprob} in terms of the number of scheduled task sets and the mean analysis time for one task set. 
%
Furthermore, comparison of four heuristics ($\mathbf{ICBFD}$, $\mathbf{ICWFD}$, $\mathbf{BCBF}$, and $\mathbf{BCWFD}$) revealed $\mathbf{ICWFD}$ as the best in terms of scheduled task sets rate.
Applying the proposed method to federated scheduling demonstrated its efficiency over the existing method~\cite{SOA-fed} for scheduling highly loaded DAG task sets unmanageable by a single core.


Future work should prioritize schedulability analysis methods that exploit DAG task parallelism. While our method improves efficiency through single-processor analysis, it remains limited to manageable DAG tasks. High-computation DAG tasks need parallel execution. Thus, high-performance methods leveraging parallelism, minimizing computation time are desirable for broader applicability.
\vspace{1mm}

\bibliography{IEEEabrv,lipics-v2021-sample-article}

\begin{thebibliography}{10}
\providecommand{\url}[1]{#1}
\csname url@samestyle\endcsname
\providecommand{\newblock}{\relax}
\providecommand{\bibinfo}[2]{#2}
\providecommand{\BIBentrySTDinterwordspacing}{\spaceskip=0pt\relax}
\providecommand{\BIBentryALTinterwordstretchfactor}{4}
\providecommand{\BIBentryALTinterwordspacing}{\spaceskip=\fontdimen2\font plus
\BIBentryALTinterwordstretchfactor\fontdimen3\font minus \fontdimen4\font\relax}
\providecommand{\BIBforeignlanguage}[2]{{%
\expandafter\ifx\csname l@#1\endcsname\relax
\typeout{** WARNING: IEEEtran.bst: No hyphenation pattern has been}%
\typeout{** loaded for the language `#1'. Using the pattern for}%
\typeout{** the default language instead.}%
\else
\language=\csname l@#1\endcsname
\fi
#2}}
\providecommand{\BIBdecl}{\relax}
\BIBdecl

\bibitem{waymo}
``Waymo,'' \url{https://waymo.com/}.

\bibitem{autoware}
S.~Kato, S.~Tokunaga, Y.~Maruyama, S.~Maeda, M.~Hirabayashi, Y.~Kitsukawa, A.~Monrroy, T.~Ando, Y.~Fujii, and T.~Azumi, ``{Autoware on Board: Enabling Autonomous Vehicles with Embedded Systems},'' in \emph{Proc. of ACM/IEEE ICCPS}, 2018, pp. 287--296.

\bibitem{AV-DAG3--DAG-WCET1}
Q.~He, J.~Sun, N.~Guan, M.~Lv, and Z.~Sun, ``{Real-Time Scheduling of Conditional DAG Tasks With Intra-Task Priority Assignment},'' \emph{IEEE Trans. Comput.-Aided Des. Integr. Circuits Syst.}, vol.~42, no.~10, pp. 3196--3209, 2023.

\bibitem{AV-DAG1}
Z.~Houssam-Eddine, N.~Capodieci, R.~Cavicchioli, G.~Lipari, and M.~Bertogna, ``{The HPC-DAG Task Model for Heterogeneous Real-Time Systems},'' \emph{IEEE Trans. Comput.}, vol.~70, no.~10, pp. 1747--1761, 2021.

\bibitem{AV-DAG2}
A.~Melani, M.~Bertogna, V.~Bonifaci, A.~Marchetti-Spaccamela, and G.~C. Buttazzo, ``{Response-Time Analysis of Conditional DAG Tasks in Multiprocessor Systems},'' in \emph{Proc. of ECRTS}, 2015, pp. 211--221.

\bibitem{DAG-WCET2}
Q.~He, N.~Guan, and M.~Lv, ``{Longer Is Shorter: Making Long Paths to Improve the Worst-Case Response Time of DAG Tasks},'' \emph{IEEE Trans. Comput.-Aided Des. Integr. Circuits Syst.}, vol.~43, no.~12, pp. 4519--4531, 2024.

\bibitem{pWCET}
S.~Bozhko, F.~Marković, G.~von~der Brüggen, and B.~B. Brandenburg, ``{What Really is pWCET? A Rigorous Axiomatic Proposal},'' in \emph{Proc. of IEEE RTSS}, 2023, pp. 13--26.

\bibitem{PTA-hybrid}
L.~BEKDEMİR and C.~F. BAZLAMAÇCI, ``{Hybrid Probabilistic Timing Analysis},'' in \emph{Proc. of UYMS}, 2021, pp. 1--6.

\bibitem{probability-timing-survey}
R.~I. Davis and L.~Cucu-Grosjean, ``{A Survey of Probabilistic Timing Analysis Techniques for Real-Time Systems},'' \emph{LITES}, pp. 1--60, 2019.

\bibitem{MBPTA2022}
J.~de~Barros~Vasconcelos and G.~Lima, ``{Possible Risks with EVT-based Timing Analysis: an Experimental Study on a Multi-core Platform},'' in \emph{Proc. of SBESC}, 2022, pp. 1--8.

\bibitem{DAGprob}
S.~Ben-amor, L.~Cucu-Grosjean, M.~Mezouak, and Y.~Sorel, ``{Probabilistic Schedulability Analysis for Precedence Constrained Tasks on Partitioned Multi-core},'' in \emph{Proc. of IEEE ETFA}, 2020, pp. 345--352.

\bibitem{SHPC-DAG}
X.~Deng, A.~H. Sifat, S.-Y. Huang, S.~Wang, J.-B. Huang, C.~Jung, R.~Williams, and H.~Zeng, ``{Partitioned Scheduling with Safety-Performance Trade-Offs in Stochastic Conditional DAG Models},'' \emph{J. Syst. Archit.}, vol. 153, p. 103189, 2024.

\bibitem{DAG-E2E}
H.~Lee, Y.~Choi, T.~Han, and K.~Kim, ``{Probabilistically Guaranteeing End-to-End Latencies in Autonomous Vehicle Computing Systems},'' \emph{IEEE Trans. Comput.}, vol.~71, no.~12, pp. 3361--3374, 2022.

\bibitem{MonteCarlo}
S.~Bozhko, G.~von~der Brüggen, and B.~B. Brandenburg, ``{Monte Carlo Response-Time Analysis},'' in \emph{Proc. of IEEE RTSS}, 2021, pp. 342--355.

\bibitem{probability-survey}
R.~I. Davis and L.~Cucu-Grosjean, ``{A Survey of Probabilistic Schedulability Analysis Techniques for Real-Time Systems},'' \emph{LITES}, vol.~6, no.~1, pp. 04:1--04:53, 2019.

\bibitem{berry-essen}
F.~Marković, T.~Nolte, and A.~V. Papadopoulos, ``{Analytical Approximations in Probabilistic Analysis of Real-Time Systems},'' in \emph{Proc. of IEEE RTSS}, 2022, pp. 158--171.

\bibitem{EDF}
G.~von~der Brüggen, N.~Piatkowski, K.-H. Chen, J.-J. Chen, K.~Morik, and B.~B. Brandenburg, ``{Efficiently Approximating the Worst-Case Deadline Failure Probability Under EDF},'' in \emph{Proc. of IEEE RTSS}, 2021, pp. 214--226.

\bibitem{2023-FP-RT}
K.~Zagalo, Y.~Abdeddaïm, A.~Bar-Hen, and L.~Cucu-Grosjean, ``{Response Time Stochastic Analysis for Fixed-Priority Stable Real-Time Systems},'' \emph{IEEE Trans. Comput.}, vol.~72, no.~1, pp. 3--14, 2023.

\bibitem{Baruah2013PartitionedEDF}
S.~K. Baruah, ``{Partitioned EDF scheduling: a closer look},'' \emph{Real-Time Syst.}, vol.~49, pp. 715--729, 2013.

\bibitem{bin-packing}
C.~Mommessin, T.~Erlebach, and N.~V. Shakhlevich, ``{Classification and Evaluation of the Algorithms for Vector Bin Packing},'' \emph{Comput. Oper. Res.}, vol. 173, p. 106860, 2025.

\bibitem{first-fed}
J.~Li, J.~J. Chen, K.~Agrawal, C.~Lu, C.~Gill, and A.~Saifullah, ``{Analysis of Federated and Global Scheduling for Parallel Real-Time Tasks},'' in \emph{Proc. of ECRTS}, 2014, pp. 85--96.

\bibitem{RD-Gen}
A.~Yano and T.~Azumi, ``{{RD-Gen}: Random {DAG} Generator Considering Multi-rate Applications for Reproducible Scheduling Evaluation},'' in \emph{Proc. of IEEE ISORC}, 2023.

\bibitem{ECRTS-toba}
H.~Toba and T.~Azumi, ``{Deadline Miss Early Detection Method for DAG Tasks Considering Variable Execution Time},'' in \emph{Proc. of ECRTS}, Dagstuhl, Germany, 2024, pp. 8:1--8:21.

\bibitem{SOA-fed}
Q.~He, N.~Guan, Z.~Jiang, and M.~Lv, ``{On the degree of parallelism for parallel real-time tasks},'' \emph{J. Syst. Archit.}, vol. 156, p. 103286, 2024.

\bibitem{light-task-fed}
S.~Baruah, ``{The federated scheduling of systems of conditional sporadic DAG tasks},'' in \emph{Proc. of EMSOFT}, 2015, pp. 1--10.

\bibitem{srv-dag}
Z.~Tong, S.~Ahmed, and J.~H. Anderson, ``{Holistically Budgeting Processing Graphs},'' in \emph{Proc. of IEEE RTSS}, 2023, pp. 27--39.

\bibitem{dag-survey}
M.~Verucchi, I.~S.~n. Olmedo, and M.~Bertogna, ``{A Survey on Real-Time DAG Scheduling, Revisiting the Global-Partitioned Infinity War},'' \emph{Real-Time Syst.}, vol.~59, no.~3, p. 479–530, 2023.

\bibitem{energy-efficient-partition}
A.~Guasque, P.~Balbastre, A.~Crespo, and J.~Coronel, ``{Energy Efficient Partition Allocation in Partitioned Systems},'' \emph{IFAC-PapersOnLine}, vol.~51, no.~10, pp. 82--87, 2018.

\bibitem{harmonic-partitioning}
T.~Wang, S.~Homsi, L.~Niu, S.~Ren, O.~Bai, G.~Quan, and M.~Qiu, ``{Harmonicity-Aware Task Partitioning for Fixed Priority Scheduling of Probabilistic Real-Time Tasks on Multi-Core Platforms},'' \emph{ACM Trans. Embed. Comput. Syst.}, vol.~16, no.~4, 2017.

\bibitem{circle-convo}
F.~Markovi\'{c}, A.~V. Papadopoulos, and T.~Nolte, ``{On the Convolution Efficiency for Probabilistic Analysis of Real-Time Systems},'' in \emph{Proc. of ECRTS}, 2021, pp. 16:1--16:22.

\bibitem{critical}
K.-H. Chen, M.~Günzel, G.~von~der Brüggen, and J.-J. Chen, ``{Critical Instant for Probabilistic Timing Guarantees: Refuted and Revisited},'' in \emph{Proc. of IEEE RTSS}, 2022, pp. 145--157.

\bibitem{srv-dag-ext}
Z.~Tong and J.~H. Anderson, ``{Budgeting Processing Graphs Under Restricted Parallelism},'' in \emph{Proc. of IEEE SIES}, 2024, pp. 172--182.

\bibitem{mixed-cri-fed}
F.~Guan, J.~Lee, C.~J. Xue, J.-M. Wu, and N.~Guan, ``{Mixed-Criticality Federated Scheduling for Relaxed-Deadline DAG Tasks},'' in \emph{Proc. of IEEE RTSS}, 2024, pp. 362--374.

\bibitem{new-fed}
F.~Guan, L.~Peng, and J.~Qiao, ``{A New Federated Scheduling Algorithm for Arbitrary-Deadline DAG Tasks},'' \emph{IEEE Trans. Comput.}, vol.~72, no.~8, pp. 2264--2277, 2023.

\bibitem{pro-fed}
J.~Li, K.~Agrawal, C.~Gill, and C.~Lu, ``{Federated scheduling for stochastic parallel real-time tasks},'' in \emph{Proc. of IEEE RTCSA}, 2014, pp. 1--10.

\end{thebibliography}

\end{document}